# A ~1$^h$ X-ray periodicity in an active galaxy RE J1034+396

Marek Gierliński[1], Matthew Middleton[1], Martin Ward[1] & Chris Done[1]

**Active galactic nuclei and quasars are thought to be scaled up versions of Galactic black hole binaries, powered by accretion onto supermassive black holes with masses of $10^6$–$10^9$ $M_\odot$, as opposed to the ~10 $M_\odot$ in binaries. One example of the similarities between these two types of systems is the characteristic rapid X-ray variability seen from the accretion flow[1]. The power spectrum of this variability in black hole binaries consists of a broad, band-limited noise with multiple quasi-periodic oscillations superimposed, where power is concentrated over a narrow range of frequencies. Although the broad noise component has been observed in many active galactic nuclei[2,3], there are no significant detections of quasi-periodic oscillations[4-6]. Here we report the discovery of a ~1$^h$ X-ray periodicity in a bright active galaxy RE J1034+396. The signal is highly statistically significant (at the 5.6σ level) and very coherent, with quality factor Q > 16. This reinforces the link between stellar and supermassive black holes, emphasizing the universal properties of accretion onto objects with very different masses. The X-ray modulation arises from the direct vicinity of the black hole, so this provides a new tool for studying active galactic nuclei.**

RE J1034+396 is a nearby ($z$ = 0.042) active galaxy, spectroscopically classified as a narrow-line Seyfert 1 (NLS1). These objects have strong emission lines produced by high density gas ionized by the UV and X-ray radiation from the accretion flow. These lines are rather narrow compared to the velocity widths seen in more typical broad line active galactic nuclei (AGN). This fact, together with other evidence, has led to the suggestion that they host supermassive black holes less massive than those inferred in a typical AGN of similar luminosity[7].

From a long (91 ks) observation using the X-ray satellite XMM-Newton we extracted a light curve for RE J1034+396 (Fig. 1), over the energy band 0.3–10 keV. Even by eye it shows an evident periodic oscillation. To more rigorously test for the presence of a periodic signal we folded the light curve with various trial periods and analysed the root-mean-square (rms) amplitude of the resulting pulse profile as a function of the period. We found a strong peak at 3730±130 s (FWHM). We used the best-fitting period to plot the expected times of minima in Fig. 1. This shows that the periodicity changes its character at around $t_0$ = 25 ks. After that time the troughs in the light curve follow the predicted minima very well, for almost 16 cycles, indicating a highly coherent signal, but before $t_0$ the troughs are shifted in phase and there are occasional additional minima. This shows that the feature is not a true periodicity, but that it wanders in phase, amplitude and/or frequency, as seen in the quasi-periodic oscillations (QPOs) in black hole binaries (BHB)[8]. Hence, we will refer to this signal as a QPO hereafter.

We concentrate first on the coherent part of the light curve (segment 2 in Fig. 1). Fig. 2 shows this light curve segment folded with the best-fitting period, while Fig 3 shows its periodogram with a strong peak at ~2.7×10$^{-4}$ Hz. In order to quantify the statistical significance of the peak we adopt the method proposed by Vaughan[9] to test the significance of periodicities against the red noise. This method involves dividing the periodogram by the best-fitting power law and using the known distribution of the periodogram ordinates to estimate the likelihood of observing a given peak. The confidence limits (3σ and 99.99%) shown in Fig. 3 are calculated including the uncertainties in the red noise model. The QPO is well above these limits, and we find that it is statistically significant at the 1 – (2×10$^{-8}$) level (~5.6σ). Even in the total light curve, including segment 1 which has less obvious periodicity, the signal is still significant at the ~3.4σ level.

This method assumed that the underlying red noise has a power-law shape but there can be breaks in this continuum, changing the derived significance of the QPO. We tested this with Monte-Carlo simulations, generating a series of light curves following a given power spectral distribution[10]. The simulated light curves had the same number of bins, mean count rate and variance, as the observed light curve. We then calculated periodograms for each of them and found the power corresponding to the upper 3σ limit in each frequency channel (the maximum significance in this method is 3.8σ due to the limited number of simulated light curves). The results for a single power-law distribution with the index taken from the best fit to our data (with index uncertainties taken into account) are comparable to the analytical limits shown in Fig. 3, as expected. A broken power law with indices -1 and -2 below and above the break frequency of $2.7\times10^{-4}$ Hz, respectively, decreases the confidence limits around the break, but the QPO remains highly significant (>3.8σ). Therefore, we conclude that the observed signal at ~$2.7\times10^{-4}$ Hz in RE J1034+396 is significant at a very high statistical level, irrespective of the assumed model for the continuum power.

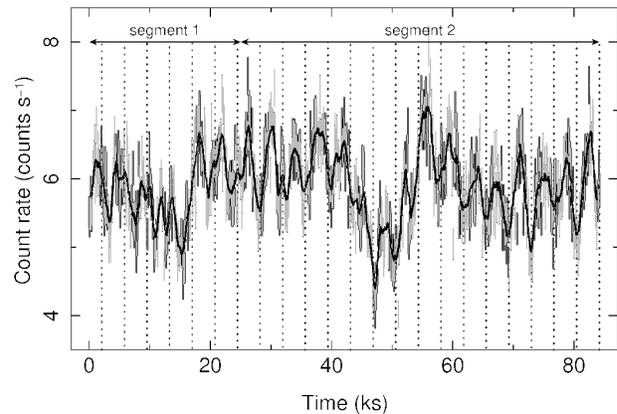

**Figure 1 | XMM-Newton light curve of RE J1034+396.** The start time of this observation was 2007-05-31 20:10:12 UTC. We extracted source and background light curves from PN, MOS1 and MOS2 cameras in 0.3–10 keV energy band, using 45 arcsec circular selection regions and rejecting the final ~7 ks due to background flares. We then combined data from all cameras together. This gave a mean count rate of 5.9±0.6 and 0.04±0.08 (s.d.) counts s$^{-1}$ for the net and background light curves, respectively. The data points, binned in 100-s intervals, are plotted in gray. The error bars are 1 s.d. The black thick curve represents the running average over 9 bins around a given bin and shows clear periodicity. The dotted vertical lines show the expected times of minima obtained from folding segment 2 with the period of 3733 s. In this paper we mainly analyse segment 2, showing a periodicity with high coherence. It contains 593 contiguous 100-s bins and almost 16 full cycles of the periodic signal. The fractional rms variability (in terms of excess variance[21]) in this segment is 9.2±0.2%.

The QPO lies within just one frequency bin of the periodogram, which makes it highly coherent, with a quality factor $Q = f/\Delta f > 16$, where $f$ and $\Delta f$ are the bin frequency and width, respectively. The rms fractional variability in the QPO is ~4.7% in the 0.3-10 keV energy band, which constitutes about half of the rms variability in the light curve. The strength of the QPO depends significantly on energy, increasing from ~2% in the 0.2–0.3 keV band, to ~10% above 1 keV. There is a time lag of ~260 s in the QPO phase between 2–10 and 0.3–0.4 keV energy bands (softer X-rays lagging behind harder X-rays).

This is the first time such a strong QPO has been convincingly found in the X-ray light curve of any AGN. Earlier claims all failed a more stringent statistical analysis[4-6], such as that used here. The only exception was that claimed for NGC6814, which turned out to be from a Galactic cataclysmic variable (CV) along the line of sight[11]! However, XMM-Newton is an imaging instrument and has a positional accuracy of ~2–3 arcsec, so considering that the number of X-ray detected CVs is about a hundred[12], the probability of a chance superposition with the AGN is vanishingly small, <$10^{-8}$. We therefore conclude that the X-ray source is associated with the galaxy. Furthermore, its X-ray luminosity ($4\times10^{43}$ erg s$^{-1}$ in 0.3–10 keV band) is then too bright for an ultra-luminous binary X-ray source, which have luminosities below ~$10^{41}$ erg s$^{-1}$. Therefore it must originate from the AGN.

The black hole mass in RE J1034+396 is not well determined, and different methods of measurement give conflicting results[13]. The virial mass derived from the Hβ emission line velocity dispersion in the broad-line region[14] is $6.3\times10^5$ M$_\odot$, and differs significantly from an estimate of $3.6\times10^7$ M$_\odot$ obtained using [OIII] as a proxy for the stellar velocity dispersion in the bulge[15]. On the other hand, if the mass-velocity dispersion relation is different in NLS1[16] than in broad-line galaxies[14], then the latter mass measurement may be overestimated. Our observed periodicity, if it is related to the Keplerian period of the innermost circular stable orbit, would correspond to the central mass between $8\times10^6$ and $9\times10^7$ $M_\odot$, for a non-rotating and maximally rotating black hole, respectively.

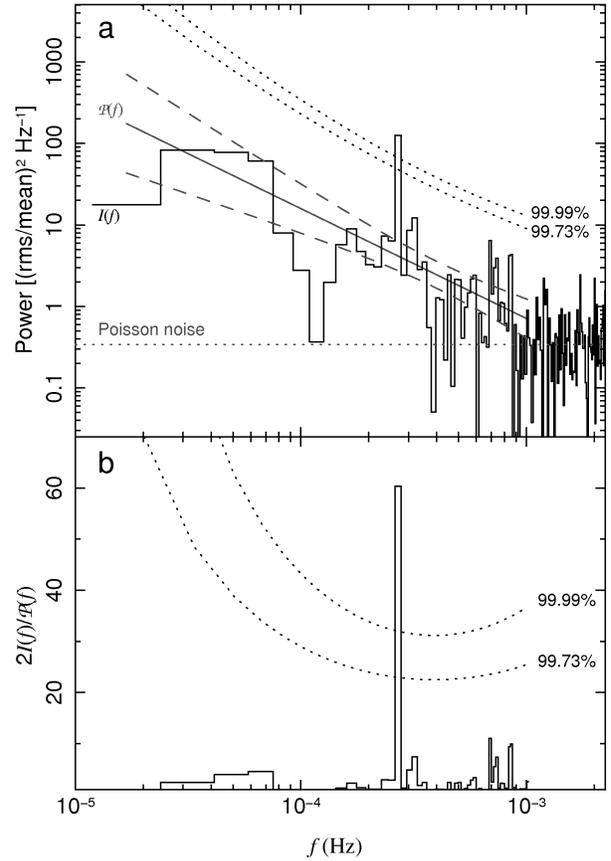

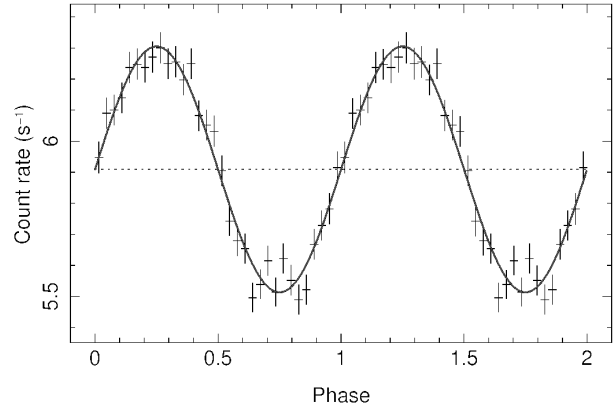

**Figure 2 | Folded light curve.** We folded the part of the light curve with significant periodicity (segment 2 in Fig. 1) with the period of 3733 s. Errors are propagated from the unfolded light curve and represent 1 s. d. Two cycles are plotted for clarity. The solid line represents the best-fitting sinusoid, the dotted line the mean count rate. The amplitude of the sinusoid is ~6.7% of the mean which corresponds to ~4.7% of the fractional rms variability in the pulse profile.

**Figure 3 | Power spectral distribution.** Panel *a* shows the power spectrum (296 data bins), $I(f)$, normalized to (rms/mean)$^2$ per Hz, calculated from segment 2 of the light curve (Fig. 1). The solid line is the de-biased[9] best-fitting (least squares method) power law, $\mathcal{P}(f)$, with index -1.35±0.18. The fit excluded the data above $10^{-3}$ Hz, dominated by the white noise. We checked that a particular choice of the cut-off did not affect the significance of the peak. The dashed curves represent the uncertainty in the power-law model. The dotted horizontal line shows the expected level of the Poissonian noise. If $\mathcal{P}(f)$ represents the true power distribution (which is our null hypothesis) then the quantity $2I(f)/\mathcal{P}(f)$ is scattered with $\chi^2$ distribution with two degrees of freedom[9,22]. The Kolmogorov-Smirnov test returned the *P* value of ~84%, so the null hypothesis is not rejected at the significance level of, e.g., α = 5%. This shows that a single power law is a good description of the underlying noise process. We also confirm this by Monte Carlo simulations, which allow us to estimate 1 s. d. errors in each frequency channel and calculate $\chi^2_\nu$ = 68.4 at 58 degrees of freedom. We use the $\chi^2$ distribution of $2I(f)/\mathcal{P}(f)$ to calculate the confidence limits on the suspected periodic signal. Panel *b* shows $2I(f)/\mathcal{P}(f)$ together with 3σ (99.73%) and 99.99% confidence limits. The same confidence limits are also plotted in panel *a*. We see that the periodic signal at ~$2.7\times10^4$ Hz is very strong, significant at a level in excess of 99.99% (the actual significance level is ~5.6σ). The significances are global, i.e. corrected for the number of frequencies tested. This result is confirmed by Monte Carlo simulations. The periodic signal is also highly significant in the individual light curves extracted from each separate X-ray camera.

Galactic BHB show a huge variety of QPOs, differing in frequency, power and coherence[17]. Generally, they can be divided into the low- and high-frequency QPOs, with frequencies <50 and >100 Hz, respectively. The strongest and most coherent low-frequency QPOs are typically seen at frequencies <10 Hz, which, when scaled to the frequency of $2.7\times10^{-4}$ Hz reported here, would imply a black hole mass of less than ~$4\times10^5$ $M_\odot$ in RE J1034+396. Given the observed bolometric luminosity (dominated by the soft X-ray/far UV component) of ~$5\times10^{44}$ erg s$^{-1}$, such a low mass requires that the system is radiating at about 10 times the Eddington limit ($L_{Edd}$).

High-frequency QPOs are occasionally seen in several BHB with high coherence[8]. They sometimes occur in pairs with the frequency ratio of 3:2[8]. These frequencies appear to be stable and are regarded as a signature of strong gravity in the vicinity of a rotating black hole[18]. A tentative frequency-mass relation, $f_0 = 931\,(M/M_\odot)^{-1}$ Hz, can be derived from three objects[8]. Here $f_0$ is the fundamental frequency of the pair, i.e. the observed frequencies are $2f_0$ and $3f_0$ (the fundamental is not seen). This relation yields the black hole mass in RE J1034+396 of $6.9\times10^6$ or $1.0\times10^7$ $M_\odot$, depending on whether the observed periodicity corresponds to $2f_0$ or $3f_0$, respectively. This would imply a luminosity of 0.5 or 0.3 $L_{Edd}$, respectively. All this clearly shows that better mass estimates (e.g. reverberation mapping or accurate stellar velocity dispersion) are required for RE J1043+396 before the QPO type can be uniquely identified.

Finally we should ask why RE J1034+396 is unique in showing the first convincing evidence for a QPO, given that many AGN have comparable quality X-ray data and have been monitored for similar or longer timescales. Perhaps we are exceptionally lucky in detecting a QPO with a small duty cycle (the high-frequency QPO in BHB is seen only occasionally). Alternatively, it may be connected to the fact that RE J1034+396 is extreme even amongst NLS1, with an unusual spectral energy distribution peaking in the far UV[19,20]. This component extends into the soft X-ray band pass, but not to the harder X-rays where the QPO is seen. Thus the far UV/soft X-ray component cannot be directly responsible for the QPO signal, but it may indicate that RE J1034+396 has an extreme mass accretion rate and that this drives both the unusual spectrum and the QPO.

QPOs remain enigmatic, but they clearly contain information about the dynamics of the infalling material. The larger mass of an AGN means that we see fewer cycles of a QPO, but with much higher time resolution compared to BHB. Therefore, future studies of such phenomena in AGN will shed new light on the origin of QPOs, and in turn, our understanding of accretion flows around black holes.

1. M$^c$Hardy, I. M., Koerding, E., Knigge, C., Uttley, P. & Fender, R. P. Active galactic nuclei as scaled-up Galactic black holes. *Nature* **444**, 730-732 (2006).
2. Edelson, R., Nandra, K. A Cutoff in the X-Ray Fluctuation Power Density Spectrum of the Seyfert 1 Galaxy NGC 3516, *Astrophys. J.* **514**, 682-690 (1999).
3. Markowitz, A., et al. X-Ray Fluctuation Power Spectral Densities of Seyfert 1 Galaxies. *Astrophys. J.* **593**, 96-114 (2003).
4. Benlloch, S., Wilms, J., Edelson, R., Yaqoob, T. & Staubert, R. Quasi-periodic Oscillation in Seyfert Galaxies: Significance Levels. The Case of Markarian 766. *Astrophys. J.* **562**, L121-124 (2001).
5. Vaughan, S., Uttley, P. Where are the X-ray quasi-periodic oscillations in active galaxies? *Mon. Not. R. Astron. Soc.* **362**, 235-244 (2005).
6. Vaughan, S. & Uttley, P. Detecting X-ray QPOs in active galaxies. *Adv. Space Res.* **38**, 1405-1408 (2006).
7. Boller, T., Brandt, W. N. & Fink, H. Soft X-ray properties of narrow-line Seyfert 1 galaxies. *Astron. Astrophys.* **305**, 53-73 (1996).
8. Remillard, R. A. & McClintock, J. E. X-Ray Properties of Black-Hole Binaries. *Ann. Rev. Astron. Astrophys.* **44**, 49-92 (2006).
9. Vaughan, S. A simple test for periodic signals in red noise. *A&A* **431**, 3n-403 (2005).
10. Timmer, J. & Koenig, M. On generating power law noise. *Astron. Astrophys.* **300**, 707-710 (1995).
11. Madejski, G. M. et al. Solving the Mystery of the X-Ray Periodicity in the Seyfert Galaxy NGC6814, *Nature* **365**, 626-628 (1993).
12. Verbunt, F., Bunk, W. H., Ritter, H. & Pfeffermann, E. Cataclysmic variables in the ROSAT PSPC All Sky Survey. *Astron. Astrophys.* **327**, 602-613 (1997).
13. Bian, W., Zhao, Y., Black hole masses in narrow-line Seyfert 1 galaxies. *Mon. Not. R. Astron. Soc.* **352**, 823-827 (2004).
14. Kaspi, S., et al. Reverberation Measurements for 17 Quasars and the Size-Mass-Luminosity Relations in Active Galactic Nuclei. *Astrophys. J.* **533**, 631-649 (2000).
15. Tremaine, S., et al. The Slope of the Black Hole Mass versus Velocity Dispersion Correlation. *Astrophys. J.* **574**, 740-753 (2002).
16. Grupe, D. & Mathur, S. $M_{BH}$-σ relation for a complete sample of soft x-ray–selected active galactic nuclei. *Astrophys. J.* **606**, L41-44 (2004).
17. Klein-Wolt, M. & van der Klis, M. Identification of Black Hole Power Spectral Components across All Canonical States. *Astrophys. J.* **675**, 1407-1423 (2008).
18. Abramowicz, M. A. & Kluźniak, W. A precise determination of black hole spin in GRO J1655-40. *Astron. Astrophys.* **374**, L19-20 (2001).
19. Pounds, K. A., Done, C. & Osborne, J. P. RE 1034+39: a high-state Seyfert galaxy? *Mon. Not. R. Astron. Soc.* **277**, L5-10 (1995).
20. Casebeer, D. A., Leighly, K. M. & Baron, E. FUSE Observation of the Narrow-Line Seyfert 1 Galaxy RE 1034+39: Dependence of Broad Emission Line Strengths on the Shape of the Photoionizing Spectrum. *Astrophys. J.* **637**, 157-182 (2006).
21. Vaughan, S., Edelson, R., Warwick, R. S. & Uttley, P. On characterizing the variability properties of X-ray light curves from active galaxies. *Mon. Not. R. Astron. Soc.* **345**, 1271-1284 (2003).
22. Papadakis, I. E. & Lawrence, A. Improved Methods for Power Spectrum Modelling of Red Noise. *Mon. Not. R. Astron. Soc.* **261**, 612-624 (1993).

**Acknowledgements** MG has been supported in part by the Polish MNiSW grant NN203065933 (2007-2010). CD acknowledges financial support through a PPARC Senior Fellowship.